\documentclass[conference]{IEEEtran}
\IEEEoverridecommandlockouts
\usepackage{cite}
\usepackage{amsmath,amssymb,amsfonts}
\usepackage{algorithmic}
\usepackage{graphicx}
\usepackage{textcomp}
\usepackage{xcolor}
\usepackage{wrapfig}
\usepackage{booktabs}
\usepackage{multirow}
\usepackage{float}
\usepackage{url}
\usepackage{hyperref}
\def\BibTeX{{\rm B\kern-.05em{\sc i\kern-.025em b}\kern-.08em
T\kern-.1667em\lower.7ex\hbox{E}\kern-.125emX}}

\newcommand{\cmark}{\checkmark}
\newcommand{\xmark}{\(\times\)}

\begin{document}

\title{Causal Feature Selection for Weather-Driven Residential Load Forecasting
}

\newcommand{\ind}{\perp\!\!\!\!\perp}

\makeatletter
\newcommand{\linebreakand}{%
\end{@IEEEauthorhalign}
\hfill\mbox{}\par
\mbox{}\hfill
\begin{@IEEEauthorhalign}
}
\makeatother

\author{%
  \IEEEauthorblockN{%
    Elise Zhang\IEEEauthorrefmark{1},
    Fran\c{c}ois Mirall\`es\IEEEauthorrefmark{2},
    St\'ephane Dellacherie\IEEEauthorrefmark{3}\IEEEauthorrefmark{4},
    Di Wu\IEEEauthorrefmark{1},
  Benoit Boulet\IEEEauthorrefmark{1}}
  \IEEEauthorblockA{\IEEEauthorrefmark{1}\footnotesize
  \textit{Department of Electrical and Computer Engineering, McGill University}, Montréal, QC, Canada}
  \IEEEauthorblockA{\IEEEauthorrefmark{2}\footnotesize
  \textit{Hydro-Québec Research Institute}, Varennes, QC, Canada}
  \IEEEauthorblockA{\IEEEauthorrefmark{3}\footnotesize
  \textit{Energy System Control Planning Division, Hydro-Québec}, Montréal, QC, Canada}
  \IEEEauthorblockA{\IEEEauthorrefmark{4}\footnotesize
  \textit{Department of Computer Science, Université du Québec à Montréal (UQÀM)}, Montréal, QC, Canada}
  \IEEEauthorblockA{\footnotesize
    Emails: elise.zhang@mail.mcgill.ca,
    \{miralles.francois,
    dellacherie.stephane\}@hydroquebec.com, \{di.wu5,
  benoit.boulet\}@mcgill.ca}
}


\maketitle

\begin{abstract}
  Weather is a dominant external driver of residential electricity demand, but adding many meteorological covariates can inflate model complexity and may even impair accuracy. Selecting appropriate exogenous features is non-trivial and calls for a principled selection framework, given the direct operational implications for day-to-day planning and reliability. This work investigates whether causal feature selection can retain the most informative weather drivers while improving parsimony and robustness for short-term load forecasting. We present a case study on Southern Ontario with two open-source datasets: (i) IESO hourly electricity consumption by Forward Sortation Areas; (ii) ERA5 weather reanalysis data. We compare different feature selection regimes (no feature selection, non-causal selection, PCMCI-causal selection) on city-level forecasting with three different time series forecasting models: GRU, TCN, PatchTST. In the feature analysis, non-causal selection prioritizes radiation and moisture variables that show correlational dependence, whereas PCMCI-causal selection emphasizes more direct thermal drivers and prunes the indirect covariates. We detail the evaluation pipeline and report diagnostics on prediction accuracy and extreme-weather robustness, positioning causal feature selection as a practical complement to modern forecasters when integrating weather into residential load forecasting.
\end{abstract}


\section{Introduction}

Short-term load forecasting (STLF) estimates near-future electricity demand (hours to days ahead) so operators can schedule generation, ensure reliability, and manage markets with minimal cost and risk. As grid-scale storage remains limited and costly, supply must closely track demand in real time; accurate STLF is therefore central to reliable and economical grid operations. At city level, STLF typically augments autoregressive load history with calendar signals (hour, weekday, holidays) and exogenous weather (temperature, humidity, cloud cover, radiative fluxes, precipitation). Prior studies~\cite{hong2016probabilistic, pinheiro2023short} find that appropriate weather integration improves forecasting performance. Yet, adding too many meteorological covariates inflates the feature space, which adds to model complexity and confounds the model with spurious correlations. This might negatively impact model performance, especially under seasonal regime shifts.

This paper studies whether \textbf{causal feature selection} (causal FS) helps isolate the most relevant weather drivers for electric load while reducing input dimensionality. We present a focused \textbf{case study on Southern Ontario}, pairing administrative-level residential electricity consumption from Independent
Electricity System Operator (IESO)'s \href{https://reports-public.ieso.ca/public/HourlyConsumptionByFSA/}{\textit{Hourly Consumption by FSA}} reports with ERA5~\cite{soci2024era5} hourly meteorological reanalysis data. We compare no selection, a non-causal filter, and PCMCI-based causal selection on single-city forecasting with GRU~\cite{chung2014empirical}, TCN~\cite{bai2018empirical}, and PatchTST~\cite{nie2022time}.

Our contributions are summarized as follows: (\textit{i}) A weather-informed load forecasting case study and evaluation pipeline that compares feature-selection regimes across forecasts on multiple cities; (\textit{ii}) A feature-level analysis showing that causal FS favors direct thermal drivers consistent with domain mechanisms, whereas non-causal filtering retains more indirect correlates; (\textit{iii}) we report diagnostics on model accuracy and robustness under extreme weather, positioning causal FS as a model-agnostic module for weather integration.


The remainder of the paper is structured as follows: Section~\ref{sec:related-work} introduces related work; Section~\ref{sec:prob-form} formulates the problem; Section~\ref{sec:method} presents the causal and non-causal selection modules and evaluation design; Section~\ref{sec:exp} presents the experimental results; and Section~\ref{sec:conclu} summarizes key takeaways and briefly discusses future work.

\section{Related Work}
\label{sec:related-work}

\subsection{Feature Selection for Load Forecasting}
STLF methods commonly combine load history with calendar features, and exogenous meteorology, reflecting the well-documented sensitivity of electricity demand to weather across seasons. Existing studies~\cite{hong2016probabilistic, pinheiro2023short} consistently report that weather is an important exogenous driver of electric load, and that appropriate weather integration improves accuracy and reliability. As the feature space grows with exogenous drivers, selecting appropriate input features becomes critical for both model generalization and efficiency. Classical feature selection (FS) seeks to identify the minimal feature set that still exhibits optimal prediction performance.~\cite{guyon2003introduction} FS approaches are often grouped into: (i) \emph{filters} (e.g., maximal relevance/minimal redundancy, mRMR, using mutual information)~\cite{peng2005feature}; (ii) \emph{wrappers}, which search subsets with a predictive model in the loop~\cite{el2016review}; and (iii) \emph{embedded} methods that shrink/select during training (e.g., LASSO/Elastic Net)~\cite{tibshirani1996regression}. While these approaches can reduce error and complexity, they primarily exploit correlations with the prediction target rather than underlying mechanisms (true causal relations), risking spurious selections and brittleness under distribution shift. Power systems studies also report that using all features is rarely optimal, reinforcing the need for principled selection~\cite{salcedo2018feature}.


\subsection{Causality and Causal Feature Selection}

\begin{wrapfigure}{R}{0.3\columnwidth}
  \par
  \vspace{-\intextsep}
  \scriptsize
  \centering
  \includegraphics[width=0.8\linewidth]{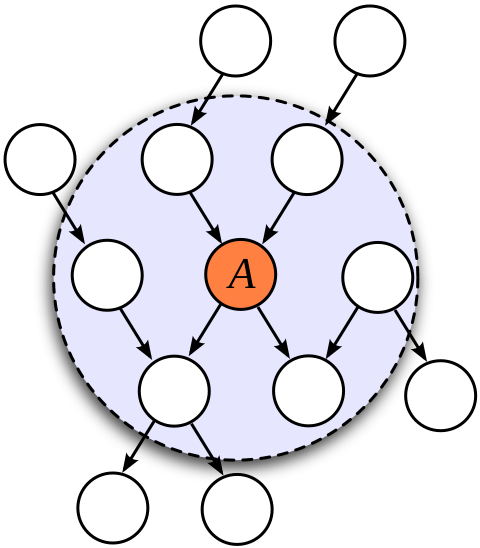}
  \caption{\scriptsize Markov Blanket: Parents, children, spouses suffice for prediction under \textit{Causal Markov} + \textit{Faithfulness}}
  \label{fig:mb}
\end{wrapfigure}

Causality captures the true data-generating mechanisms of systems, not mere predictability. Causal FS aims to select features by formal causal analysis~\cite{aliferis2010local} and seek variables that are \textbf{causally sufficient} to explain the target variable, rather than merely correlated with it. Under the Causal Markov and the Faithfulness Assumptions~\cite{hausman1999independence, spirtes2000causation}, a causal Directed Acyclic Graph (DAG) entails that each node $X$ is conditionally independent of all its non-descendants, given its direct parents. In other words, the minimal subset of features that renders $X$ independent of all remaining variable are those with a direct link with $X$ (parents, children, and ``spouses''). This subset is referred to as the \textbf{Markov blanket} (MB) of $X$ (Fig. \ref{fig:mb}), and will be sufficient to explain $X$. Any feature outside of MB is redundant for predicting $X$, given MB. Consequently, selecting MB provides a minimal and causally sufficient feature set, aligning with the goal of FS. Classical work applies this principle directly to FS: KS algorithm~\cite{koller1996toward}, which uses MB-based screening, was established as a criterion for optimal feature selection as early as 1996; subsequent work on MB induction (e.g., IAMB~\cite{tsamardinos2003algorithms}) operationalizes the discovery of MBs around a target at scale. For \textbf{multivariate time series}, the relevant blanket narrows to direct causal parents, both lagged and instantaneous, available during inference. Time series causal discovery approaches (e.g., TiMINo~\cite{peters2013causal}, PCMCI/PCMCI+~\cite{runge2019detecting, runge2020discovering}) target exactly the identification of such parents, and provide a causality-aware lens for FS tailored to time series-related tasks.


\section{Problem Formulation}
\label{sec:prob-form}

We consider hourly residential electricity demand for a set of geographically proximate cities/regions $\mathcal{C}$. For each city/region $c\in\mathcal{C}$, we observe historical load $\mathcal{\textbf{Y}}^{(c)}_{0:t}=[y^{(c)}_0, y^{(c)}_1, y^{(c)}_2, ..., y^{(c)}_t]$. Calendar-datetime features $\mathcal{\textbf{D}}_{0:t}$ are also extracted (e.g., hour-of-day, day-of-week, month, day-of-year, holidays), as well as corresponding meteorological predictors $\mathcal{\textbf{W}}^{(c)}_{0:t}$ (e.g., temperature, cloud cover, radiation, precipitation, etc.). Here, weather is treated as a causal driver of electricity demand and is assumed available at prediction time via observations.

\subsection{City-Level Short-Term Load Forecasting}
\label{sec:prob-lf}
Let $\mathcal{P}^{(c)}=[\mathbf{D},\,\mathbf{Y}^{(c)},\,\mathbf{W}^{(c)}]$ denote the full predictor set. Given a lookback window $L$ and forecast horizon $H$, we pose forecasting as supervised learning from historical predictors to future demand. In this study we use \emph{one week} lookback and \emph{one day} horizon, i.e., $L{=}168$ hours and $H{=}24$ hours. For city $c$, define its predictor history $\mathcal{P}^{(c)}_{\mathrm{hist}}=[\mathbf{D}_{(t-L+1):t},\,\mathbf{Y}^{(c)}_{(t-L+1):t},\,\mathbf{W}^{(c)}_{(t-L+1):t}]$ and target $\mathbf{Y}^{(c)}_{\mathrm{tgt}}=\mathbf{Y}^{(c)}_{(t+1):(t+H)}$. The task is, for each $c$, to learn a function $f^{(c)}:\mathcal{P}^{(c)}_{\mathrm{hist}}\mapsto \mathbf{Y}^{(c)}_{\mathrm{tgt}}$ that predicts the future load values.

\subsection{Feature Selection}
\label{sec:exp-fs}
A central question in this case study is whether \textbf{causal} feature selection, as a principled selection method, improves forecasting quality and robustness, relative to using all available predictors without selection, or to non-causal selection. We propose the following working hypotheses: \textbf{H1 (Parsimony without loss):} Causally selected feature subsets $\mathcal{\mathbf{P}}^{(\mathcal{C})}_{hist}$ enable similar or lower error with reduced input dimensionality; \textbf{H2 (Robustness):} Models trained on causally selected subsets are more robust under extreme weather conditions.




\section{Methodology}
\label{sec:method}

\subsection{PCMCI Causal Discovery for Feature Selection}

PCMCI (Peter–Clark \textit{PC algorithm} with Momentary Conditional Independence)~\cite{runge2019detecting} is a causal discovery method tailored to highly-interdependent multivariate time series. PCMCI improves upon the classic PC algorithm~\cite{spirtes1991algorithm,spirtes2000causation} via a two-phase framework: \textbf{PC-Style Condition Selection}: Starting from a fully connected time-lagged graph with lag $0\leq \tau \leq \tau_{max}$ (we set $\tau_{max}=5$ to capture short-term weather impact while keeping the search space tractable), PCMCI iteratively prunes some links through PC-style conditional independence tests. The output is a candidate set of time-lagged causal parents $pa(X_t^{(j)})$ for $X_t^{(j)}$, variable $j$ at timestamp $t$. \textbf{Momentary Conditional Independence (MCI)}: For each remaining link $X_{t-\tau}^{(i)}\rightarrow X_t^{(j)}$ yielded from phase 1, the MCI test further evaluates the dependence of $X_t^{(j)}$ on $X_{t-\tau}^{(i)}$, conditioned on the candidate parents of both nodes $X_t^{(j)}$ and $X_{t-\tau}^{(i)}$ (excluding the nodes themselves). Specifically, it tests the following dependence hypothesis:
\begin{equation}
  X_t^{(j)} \not \ind X_{t-\tau}^{(i)} \mid \operatorname{Pa}\left(X_t^{(j)}\right) \backslash\left\{X_{t-\tau}^{(i)}\right\}, \operatorname{Pa}\left(X_{t-\tau}^{(i)}\right)
\end{equation}
If after conditioning on the parent sets, this dependence is no longer significant, the link $X_{t-\tau}^{(i)}\rightarrow X_t^{(j)}$ is pruned. Conditioning on parents of both nodes increases effect size and power under autocorrelation and helps rule out indirect or spurious links.

PCMCI thus estimates a time-lagged causal graph; our \textbf{causal feature set} is defined as the putative \emph{direct lagged parents of the target} on this graph under PCMCI's assumption. For discovery of contemporaneous causal effects, PCMCI+\cite{runge2020discovering} can be used. For this study, we restrict selection to lagged parents available at prediction time, and use the open-source implementation of PCMCI from the \texttt{tigramite} library.

\subsection{Non-Causal Feature Selection Baseline}
We adopt a simple filter that (i) ranks predictors by their \emph{Mutual Information} (MI) scores with the target, and (ii) removes near-duplicates via a correlation screen. MI is chosen as it captures both linear and nonlinear association. Concretely, we compute MI using \texttt{sklearn}'s estimator, and retain predictors whose MI exceeds a data-inspected  $MI_{thres}=0.025$, chosen empirically around the elbow of the MI distributions across cities/regions (we chose this simple heuristic, as our goal is to represent a reasonable non-causal filter rather than an aggressively tuned baseline). To limit redundancy and multicollinearity, we discard candidate features whose absolute Pearson correlation with an already kept feature exceeds $|\rho|=0.8$, chosen according to the \textit{Variance Inflation Factor (VIF)} multicollinearity guideline $VIF\leq 5$; $|\rho|=0.8$ implies $VIF=\frac{1}{1-\rho^2}\approx 2.78$ which is well below the guideline.


\subsection{Evaluation Pipeline}
\label{sec:eval_pip}
We evaluate if feature selection improves city-level STLF using different models, each under controlled feature sets.

\subsubsection{Data Split}
We use sliding-window \emph{time-series cross-validation}~\cite{bergmeir2012use}: each test block is preceded by a fixed training window and an inner validation slice for early stopping; performance is averaged across folds. Scalers are fitted on training data only and applied to validation and test. (In our experiments, each sliding window spans 2 years, and we have 6 folds in total.)

\subsubsection{FS and STLF Setup}
For each city $c$, we train an individual model to predict that city’s load using a 1-week history and a 24-hour horizon. We benchmark three families—recurrent (GRU), convolutional (TCN), and attention-based (PatchTST)—to cover complementary inductive biases for sequence modeling (recurrence, convolution, attention). These backbones are widely adopted and have stable open-source implementations. Together they span the design space most commonly used in short-term load forecasting. We compare four feature regimes:
\textbf{$\mathbf{F_{0}}$ (Electricity-only):} Calendar/time features and load history from IESO; \textbf{$\mathbf{F_{1}}$ (All):} $F_{0}$ plus all ERA5 weather features (no selection); \textbf{$\mathbf{F_{2}}$ (Non-causal):} $F_{0}$ plus subset of ERA5 weather variables selected by the non-causal filter; \textbf{$\mathbf{F_{3}}$ (Causal):} $F_{0}$ plus subset of ERA5 weather variables selected by PCMCI algorithm, note that here we interpret $F_{3}$ as putative direct lagged parents of electricity demand under PCMCI’s assumptions (\textit{Causal Markov}, \textit{Faithfulness}, and \textit{no unobserved confounding}), not proven causal effects.

\subsubsection{Diagnostics}
We report: \textbf{Accuracy Across Feature Regimes:} Mean Absolute Error (MAE) and Mean Absolute Percentage Error (MAPE), averaged over rolling-origin folds, reported across model classes and feature regimes; \textbf{Out-of-Distribution (OOD) Weather Robustness:} Performance is evaluated on extreme quantiles (e.g., 5$^{th}$/95$^{th}$) of key weather drivers (temperature, precipitation). Models trained with causality-aligned features are expected to be more robust under such scenarios.

\begin{table*}[htp]
  \centering
  \caption{City-wise MAE (MWh) and MAPE (\%) for GRU, TCN, and PatchTST under four feature regime. “Top MAE/MAPE” counts the number of cities where a regime attains the best score for the given model.
  }
  \label{tab:city_results}
  \setlength{\tabcolsep}{3pt}
  \small
  \resizebox{1\textwidth}{!}{%
    \begin{tabular}{lc *{9}{cc} cc}
      \toprule
      & &
      \multicolumn{2}{c}{Toronto} &
      \multicolumn{2}{c}{Peel} &
      \multicolumn{2}{c}{Hamilton} &
      \multicolumn{2}{c}{Brantford} &
      \multicolumn{2}{c}{Waterloo} &
      \multicolumn{2}{c}{London} &
      \multicolumn{2}{c}{Oshawa} &
      \multicolumn{2}{c}{Kingston} &
      \multicolumn{2}{c}{Ottawa} &
      \multicolumn{2}{c}{Count} \\
      \cmidrule(lr){3-4}\cmidrule(lr){5-6}\cmidrule(lr){7-8}\cmidrule(lr){9-10}%
      \cmidrule(lr){11-12}\cmidrule(lr){13-14}\cmidrule(lr){15-16}\cmidrule(lr){17-18}%
      \cmidrule(lr){19-20}\cmidrule(lr){21-22}
      \textbf{Model} & \textbf{Feature Set} &
      \textbf{MAE} & \textbf{MAPE} &
      \textbf{MAE} & \textbf{MAPE} &
      \textbf{MAE} & \textbf{MAPE} &
      \textbf{MAE} & \textbf{MAPE} &
      \textbf{MAE} & \textbf{MAPE} &
      \textbf{MAE} & \textbf{MAPE} &
      \textbf{MAE} & \textbf{MAPE} &
      \textbf{MAE} & \textbf{MAPE} &
      \textbf{MAE} & \textbf{MAPE} &
      \textbf{Top MAE} & \textbf{Top MAPE} \\
      \midrule
      \multirow{4}{*}{GRU}
      & $F_{0}$ & 29.21 & 5.06 & 16.81 & 5.28 & 7.55 & 5.15 & 2.18 & 5.44 & 7.71 & 4.97 & 7.79 & 6.19 & 4.21 & 6.80 & 3.24 & 6.78 & 22.33 & 6.91 & 0 & 0 \\
      & $F_{1}$ & 29.90 & 5.12 & 17.12 & 5.45 & 8.12 & 5.29 & 2.22 & 5.37 & 8.17 & 5.17 & 7.80 & 6.16 & 4.19 & 6.63 & \textbf{3.14} & \textbf{6.53} & 23.98 & 7.11 & 1 & 1 \\
      & $F_{2}$ & 29.43 & 5.03 & 16.75 & 5.16 & \textbf{7.50} & \textbf{5.05} & \textbf{2.16} & \textbf{5.26} & 7.76 & 4.93 & 7.11 & 5.85 & 3.91 & 6.35 & 3.25 & 6.74 & 22.76 & 6.80 & 2 & 2 \\
      & $F_{3}$ & \textbf{29.17} & \textbf{5.01} & \textbf{15.82} & \textbf{4.89} & \textbf{7.50} & \textbf{5.05} & 2.20 & 5.32 & \textbf{7.51} & \textbf{4.76} & \textbf{6.84} & \textbf{5.34} & \textbf{3.84} & \textbf{6.16} & 3.23 & 6.64 & \textbf{21.27} & \textbf{6.53} & \textbf{7} & \textbf{7} \\
      \midrule
      \multirow{4}{*}{TCN}
      & $F_{0}$ & 27.06 & 4.75 & 17.60 & 5.56 & 8.25 & 5.59 & 2.19 & 5.51 & 8.06 & 5.27 & 5.97 & 4.65 & 3.64 & 6.11 & 3.81 & 7.43 & 24.00 & 7.31 & 0 & 0 \\
      & $F_{1}$ & 26.88 & 4.64 & \textbf{17.08} & \textbf{5.27} & 8.63 & 5.74 & 2.28 & 5.67 & 8.49 & 5.40 & 5.94 & 4.51 & 3.68 & 6.07 & 3.61 & 7.19 & 23.60 & 7.27 & 1 & 1 \\
      & $F_{2}$ & \textbf{26.40} & \textbf{4.55} & 17.70 & 5.55 & 8.74 & 6.00 & \textbf{2.00} & \textbf{4.96} & \textbf{7.58} & \textbf{4.89} & 5.98 & 4.52 & 3.53 & 5.86 & 4.12 & 8.17 & \textbf{22.20} & \textbf{6.79} & \textbf{4} & \textbf{4} \\
      & $F_{3}$ & 26.53 & 4.60 & 17.11 & 5.34 & \textbf{8.47} & \textbf{5.72} & 2.13 & 5.20 & 7.82 & 5.18 & \textbf{5.83} & \textbf{4.49} & \textbf{3.25} & \textbf{5.47} & \textbf{3.55} & \textbf{7.04} & 25.10 & 7.70 & \textbf{4} & \textbf{4} \\
      \midrule
      \multirow{4}{*}{PatchTST}
      & $F_{0}$ & 26.32 & 4.55 & 14.20 & 4.56 & 6.95 & 4.85 & 1.83 & 4.75 & 7.48 & 4.88 & \textbf{5.45} & 4.35 & 2.67 & 4.64 & 2.49 & 5.07 & 15.50 & 5.05 & 1 & 0 \\
      & $F_{1}$ & 25.40 & 4.44 & 15.73 & 5.08 & 7.63 & 5.26 & \textbf{1.81} & \textbf{4.63} & \textbf{7.03} & \textbf{4.54} & 5.90 & 4.69 & 2.74 & 4.69 & 2.32 & 4.70 & 15.40 & 4.99 & 2 & 2 \\
      & $F_{2}$ & 24.53 & 4.33 & \textbf{14.15} & \textbf{4.53} & 7.06 & 4.89 & 1.88 & 4.84 & 7.31 & 4.77 & 5.72 & 4.53 & 2.72 & 4.56 & 2.54 & 5.21 & \textbf{14.90} & \textbf{4.91} & 2 & 2 \\
      & $F_{3}$ & \textbf{24.37} & \textbf{4.28} & 14.78 & 4.58 & \textbf{6.91} & \textbf{4.70} & 1.85 & 4.76 & 7.18 & 4.69 & \textbf{5.45} & \textbf{4.26} & \textbf{2.59} & \textbf{4.38} & \textbf{2.31} & \textbf{4.62} & 15.40 & 5.06 & \textbf{5} & \textbf{5} \\
      \bottomrule
  \end{tabular}}
\end{table*}

\section{Experiments}
\label{sec:exp}

\begin{figure}[t]
  \centering
  \includegraphics[width=0.4\textwidth]{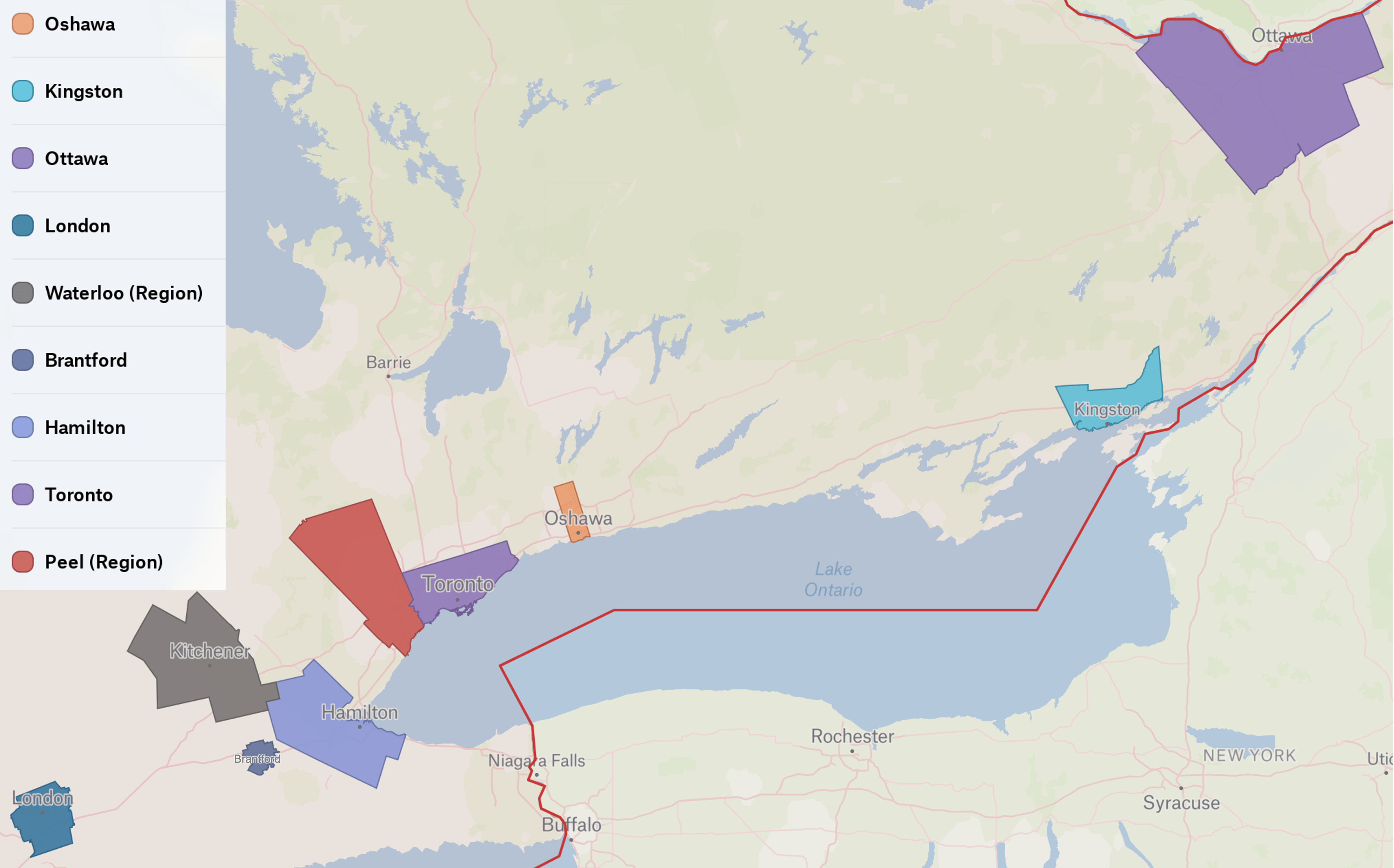}
  \caption{Administrative regions of Southern Ontario from the census divisions of the \href{https://www12.statcan.gc.ca/census-recensement/2021/geo/sip-pis/boundary-limites/index2021-eng.cfm?year=21}{\textit{2021 Census}} by \textit{Statistics Canada}. Map created with \href{https://felt.com/map}{Felt} GIS platform.}
  \label{fig:map}
\end{figure}

\subsection{Datasets}
We pair administrative-level electricity consumption with physically consistent hourly weather reanalysis for Ontario.

\subsubsection{IESO Ontario Hourly Electricity Consumption}
We use the Independent Electricity System Operator (IESO)'s public report, \href{https://reports-public.ieso.ca/public/HourlyConsumptionByFSA/}{“Hourly Consumption by Forward Sortation Area (FSA)”}. We use records from January 2018 to March 2024, and have the following fields: \texttt{FSA} (first 3 characters of a Canadian postal code), \texttt{timestamp} (hour-ending), \texttt{consumer type} (\textit{Residential} or \textit{Small General Service}), \texttt{premise count} (total number of end users recorded in the past hour period), \texttt{total consumption} (kWh). In this work, we restrict to the \textit{Residential} sector, as residential load reflects human activities and is most directly impacted by the weather conditions. Data entries are aggregated by FSA to the municipal levels to form city-scale residential load series. We choose regions of varying sizes and population levels, including: Toronto, Peel Region (Mississauga, Brampton), Hamilton, Brantford, Waterloo Region (Cambridge, Kitchener, Waterloo), London, Oshawa, Kingston, Ottawa. Geographical information of each city/region (longitudes, latitudes) is subsequently obtained to facilitate ERA5 data extraction.

\subsubsection{ERA5 Reanalysis}
ERA5~\cite{soci2024era5} is the 5$^{th}$-generation global weather reanalysis data offered by Copernicus Climate Change Service (C3S), providing hourly estimates of a lot of atmospheric, land, and ocean variables. In this work, we use \textit{ERA5 Single Level}~\cite{hersbach2023era5} and \textit{ERA5-Land}~\cite{munoz2019era5} datasets. The following covariates are extracted as potential causal drivers for analysis: total cloud cover (\texttt{tcc}), total column water (\texttt{tcw}), Earth surface temperature (\texttt{skt}), net terrestrial radiation flux (\texttt{avg-snlwrf}), net solar radiation flux (\texttt{avg-snswrf}), 2-meter air temperature (\texttt{t2m}), 2-meter dewpoint temperature (\texttt{d2m}), and total precipitation (\texttt{tp}). We use ERA5 as retrospective exogenous inputs to isolate the effect of feature selection; note that in operations, these would be replaced by meteorology forecast and absolute errors may increase.

\subsection{Feature Selection Results}

Following Section~\ref{sec:exp-fs}, our feature selection experiments yield feature subsets as in Table~\ref{tab:feature_sets}. The datetime features shared across all subsets are hour-of-day, day-of-week, month, day-of-year, and an Ontario holiday flag.
In our implementation, cyclic fields (hour, day, month) are represented with sine/cosine pairs to avoid artificial discontinuities at wrap-around boundaries. We also include \texttt{premise count} from IESO data as a slow-moving proxy for the total number of customers in each region.




\begin{table}[t]
  \centering
  \scriptsize
  \caption{Feature subsets.}
  \label{tab:feature_sets}
  \begin{tabular}{lcccc}
    \toprule
    \textbf{Feature} & \textbf{$F_{0}$} & \textbf{$F_{1}$} & \textbf{$F_{2}$} & \textbf{$F_{3}$} \\
    \midrule
    Load history            & \cmark & \cmark & \cmark & \cmark \\
    Datetime              & \cmark & \cmark & \cmark & \cmark \\
    Premise Count              & \cmark & \cmark & \cmark & \cmark \\
    \midrule
    Total Cloud Cover (\texttt{tcc})       & \xmark & \cmark & \xmark & \xmark \\
    Total Column Water (\texttt{tcw})      & \xmark & \cmark & \cmark & \xmark \\
    Earth Surface Temperature (\texttt{skt})          & \xmark & \cmark & \xmark & \cmark \\
    Net Long-Wave Radiation Flux (\texttt{avg-snlwrf})     & \xmark & \cmark & \cmark & \xmark \\
    Net Short-Wave Radiation Flux (\texttt{avg-snswrf})    & \xmark & \cmark & \cmark & \xmark \\
    2-Meter Air Temperature (\texttt{t2m})     & \xmark & \cmark & \cmark & \cmark \\
    2-Meter Dewpoint Temperature \texttt{d2m}          & \xmark & \cmark & \xmark & \xmark \\
    Total Precipitation (\texttt{tp})      & \xmark & \cmark & \cmark & \cmark \\
    \bottomrule
  \end{tabular}
  \begin{minipage}{\columnwidth}
    \footnotesize

  \end{minipage}
\end{table}

\textbf{Compare $F_{2}$ and $F_{3}$}: Both $F_{2}$ and $F_{3}$ select \texttt{t2m} (2m air temperature) and \texttt{tp} (precipitation). This is anticipated given the well-documented link, temperature$\rightarrow$load
, and also the role of rain or snow as proxies for weather regimes that alter occupancy and HVAC usage.
Their selections diverge on radiation and moisture variables: $F_{3}$ drops \texttt{tcc} and \texttt{tcw} (cloud, column water) and retains \texttt{skt} (surface temperature); while $F_{2}$ keeps \texttt{tcw} (column water), \texttt{avg-snlwrf}, and \texttt{avg-snswrf} (long- and short-wave radiation flux). As a causal discovery method, PCMCI identifies variables that potentially have a more direct effect on load and prunes features whose influence is indirect and mediated (e.g., cloud$\rightarrow$radiation$\rightarrow$temperature$\rightarrow$load, or moisture$\rightarrow$thermal comfort$\rightarrow$temperature proxies$\rightarrow$load). In this case study, radiation (\texttt{avg-snlwrf}, \texttt{avg-snswrf}) and column water (\texttt{tcw}) lose significance once temperature variables are identified as the putative direct driver of load. In short, $F_{2}$ reflects a more correlational association: it keeps radiation and column water because they explain a part of the variance of the prediction target. $F_{3}$ reflects structural parsimony: it favors more direct thermal drivers and discards variables whose effects are indirect or redundant after conditioning.

\subsection{City-Level Forecasting Results}

We train GRU, TCN, and PatchTST backbones with comparable capacity across cities.
~\footnote{\tiny L2 loss with Adam ($\text{lr}=10^{-4}$, $\beta_1=0.9$, $\beta_2=0.999$), batch size $64$, max $500$ epochs, early stopping (patience $20$, $\Delta=10^{-4}$) on validation MAE, and dropout $0.1$ where applicable. Lookback and horizon follow Sec.~\ref{sec:prob-lf} ($L{=}168$\,h, $H{=}24$\,h).
  \emph{GRU}: hidden size $d{=}64$, $4$ stacked layers, default PyTorch gating and initialization. \emph{TCN}: $4$ temporal blocks with dilation base $2$ (dilations $1,2,4,8$), kernel size $3$, $64$ channels per layer, residual connections.
  \emph{PatchTST}: encoder dimension $d_\text{model}{=}64$, $4$ Transformer encoder layers, $4$ attention heads, patch length $16$, stride $8$, standard positional encoding and pre-norm.
}
Table~\ref{tab:city_results} reports MAE (MWh) and MAPE (\%) for each city under the four feature regimes ($F_{0}$–$F_{3}$). For every backbone, $F_{3}$ attains the most city-wise best scores (Count of Top MAE/MAPE), indicating systematic gains from pruning indirect meteorological covariates. The results also confirm that using all features is not always the best: $F_{1}$ overall underperforms $F_{3}$ and $F_{2}$, suggesting that naively adding all weather variables may introduce spurious correlations and harm performance. Across architectures, selecting putative \emph{direct} drivers ($F_{3}$) generally yields modest but reliable accuracy improvements over $F_{0}$/$F_{1}$, and often outperforms a non-causal filter ($F_{2}$). Overall, $F_{3}$ tends to win across recurrent, convolutional, and Transformer-style models, supporting the view that causal FS is a model-agnostic way to integrate weather while controlling feature-set complexity.

\subsection{\textit{Out-of-Distribution} (OOD) Weather Inference Results}

\begin{figure}[t]
  \centering
  \includegraphics[width=0.8\linewidth]{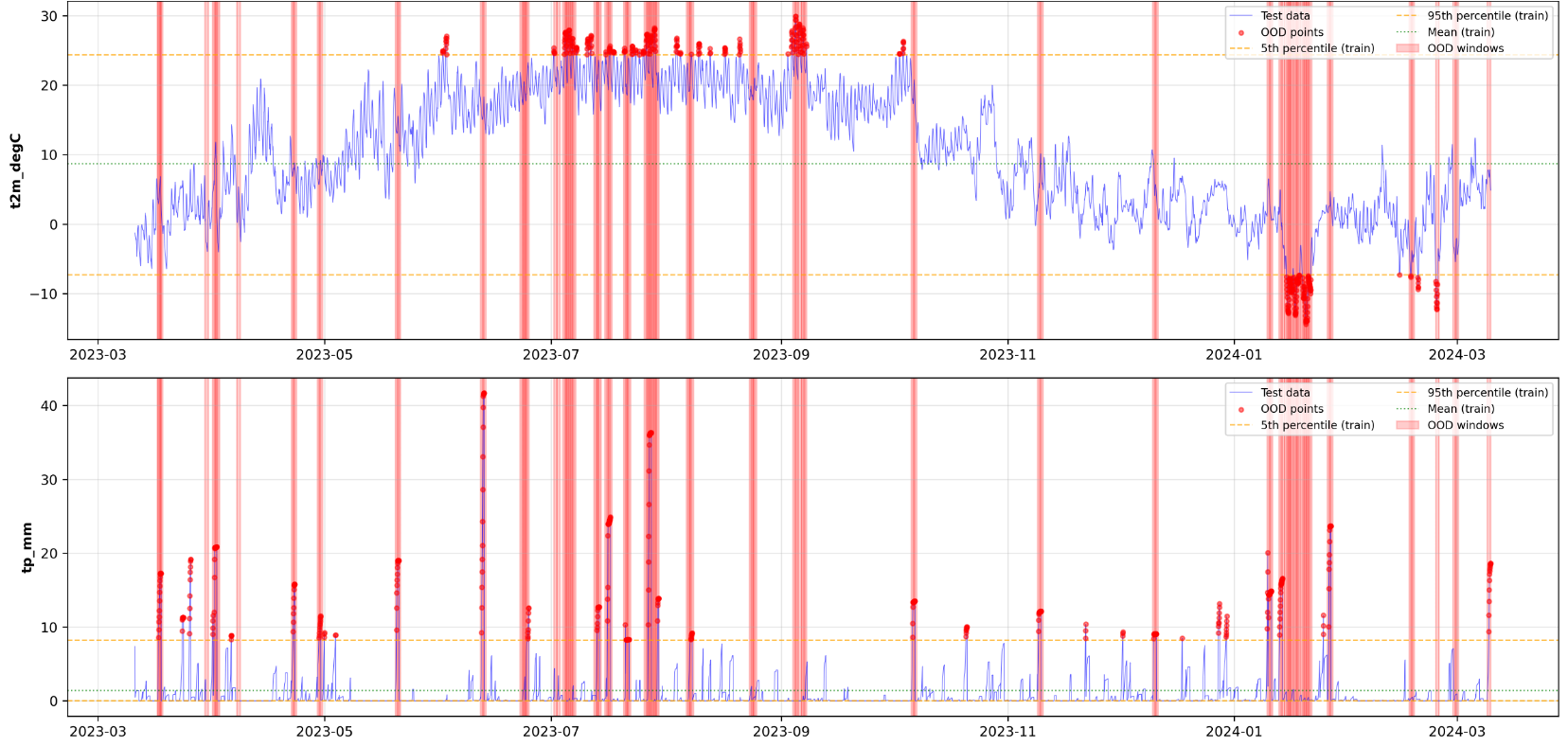}
  \caption{\scriptsize Demo: Toronto's OOD Weather Events Identified in Test Year, Including Summer Heat and Winter Cold Wave, Heavy Precipitation}
  \label{fig:ood}
\end{figure}

\begin{table}[t]
  \centering
  \scriptsize
  \caption{\scriptsize OOD-weather evaluation. Metrics: MAE (MWh) and MAPE (\%). Bold: the best for each city across all configurations. Underline: 2$^{nd}$ best. Also compute relative error reduction (\%) from 2$^{nd}$ best to the best.}
  \label{tab:ood_city_results}
  \setlength{\tabcolsep}{6pt}
  \resizebox{0.8\linewidth}{!}{%
    \begin{tabular}{lccccc}
      \toprule
      & & \multicolumn{2}{c}{\textbf{Toronto}} & \multicolumn{2}{c}{\textbf{Ottawa}} \\
      \cmidrule(lr){3-4} \cmidrule(lr){5-6}
      \textbf{Model} & \textbf{Feature Set} & \textbf{MAE} & \textbf{MAPE} & \textbf{MAE} & \textbf{MAPE} \\
      \midrule
      \multirow{4}{*}{GRU}
      & $F_{0}$ & 45.62 & 6.43 & 38.25 & 9.70 \\
      & $F_{1}$ & 48.35  & 6.87 & 27.23 & 7.40 \\
      & $F_{2}$ & 45.46 & 6.48 & 34.13 & 8.93 \\
      & $F_{3}$ & 47.65  & 6.66 & 38.89 & 9.74 \\
      \midrule
      \multirow{4}{*}{TCN}
      & $F_{0}$ & 44.73  & 6.24 & 29.19 & 7.31 \\
      & $F_{1}$ & 45.15  & 6.30 & 26.48 & 6.83 \\
      & $F_{2}$ & 43.53  & \underline{6.10} & 26.10 & 6.89 \\
      & $F_{3}$ & 44.88  & 6.31 & 27.12 & 7.05 \\
      \midrule
      \multirow{4}{*}{\textbf{PatchTST}}
      & $F_{0}$ & \underline{42.13} & \textbf{5.70} & 26.13 & 6.73 \\
      & $F_{1}$ & 47.79  & 6.84 & \underline{24.80} & \underline{6.60} \\
      & $F_{2}$ & 47.81  & 6.64 & 26.41 & 7.13 \\
      & \textbf{$F_{3}$} & \textbf{40.13} & \textbf{5.70} & \textbf{23.69} & \textbf{6.24} \\
      \midrule
      \multicolumn{2}{c}{\textit{Err. Reduction (\%)}} & \textit{4.75} & \textit{6.56} & \textit{4.48} & \textit{5.45} \\
      \bottomrule
    \end{tabular}%
  }
\end{table}

We define OOD weather relative to the history for each city: compute the 5$^{th}$/95$^{th}$ percentiles of hourly temperature (\texttt{t2m}) and precipitation (\texttt{tp}) using new training window, 2018-01-01 to 2023-03-10. Any 24h window in the held-out test year (2023-03-11 to 2024-03-10) is flagged OOD if over 50\% of its 24 hours fall outside those thresholds, with a minimum 24 h between windows. We report OOD windows for Toronto and Ottawa (E.g., Toronto OODs in Fig.~\ref{fig:ood}) in the test year. Each forecaster is retrained from scratch on the full training window with the same hyperparameters as in the city-level study. We evaluate models on the detected OOD windows across feature regimes and report average MAE and MAPE across all OOD windows (results in Table~\ref{tab:ood_city_results}). Overall, the best configuration in both cities is PatchTST with $F_{3}$ (PCMCI), outperforming all other configurations. This pattern suggests that pruning indirect or redundant weather covariates via causal selection yields a more compact and robust forecast under extreme weather conditions, particularly for attention-based models that benefit from reduced input redundancy. On the other hand, GRU and TCN show mixed behavior, occasionally favoring $F_{0}$/$F_{2}$, suggesting architecture-specific interactions with feature sparsity and redundancy that merit further study. Still, across cities and different weather OODs, PatchTST with feature set $F_{3}$ is consistently the strongest, supporting our claim that causal FS improves robustness under distribution shift while maintaining accuracy.

\section{Conclusion}
\label{sec:conclu}
This work is a dedicated case study evaluating whether causal feature selection can make weather-informed STLF more compact and robust. Across 9 Southern Ontario cities/regions, and 3 baselines (GRU, TCN and PatchTST), selecting features via PCMCI (feature set F3) generally yields a compact input feature set, matches or improves accuracy, and shows robustness under extreme cold or hot weather and heavy precipitation events. Overall, the feature selection guided by PCMCI causal discovery method can be a model-agnostic, lightweight module that improves forecasting performance while strengthening generalization under distribution shifts. Future work will expand to multi-city multivariate forecasting, causal transfers across regions, and introducing contemporaneous effects via PCMCI+ to further stress-test causal feature selection in operational settings.

\bibliographystyle{IEEEtran}
\bibliography{ref}

\end{document}